# Double Dome and Reemergence of Superconductivity in Pristine 6$R$-TaS$_2$ under Pressure

Xindeng Lv[1], Hao Song[1], Kun Chen[1], Sirui Liu[1], Yanping Huang[1,] *, Yuqiang Fang[3,] * and Tian Cui[1,2,] *

[1] Institute of High Pressure Physics, School of Physical Science and Technology, Ningbo University, Ningbo 315211, China

[2] State Key Laboratory of Superhard Materials, College of Physics, Jilin University, Changchun 130012, China

[3] State Key Laboratory of High Performance Ceramics and Superfine Microstructure, Shanghai Institute of Ceramics, Chinese Academy of Sciences, Shanghai, 200050, China

*Corresponding authors' e-mails: huangyanping@nbu.edu.cn; fangyuqiang@mail.sic.ac.cn; cuitian@nbu.edu.cn

## Abstract

Investigating the implications of interlayer coupling on superconductivity is essential for comprehending the intrinsic mechanisms of high temperature superconductors. Van der Waals heterojunctions have attracted extensive research due to their exotic interlayer coupling. Here, we present a natural heterojunction superconductor of 6$R$-TaS$_2$ that demonstrates a double-dome of superconductivity, in addition to, the reemergence of superconducting under high pressures. Our first principles calculation shows that the first dome of superconductivity in 6$R$-TaS$_2$ can be attributed to changes in interlayer coupling and charge transfer. The second superconducting dome and the reemergence of superconductivity can be ascribed to changes in the density of states resulting from Fermi surface reconstruction, in which the DOS of $T$-layer and S $p$-orbitals play a crucial role. We have reported the first observation in TMDs that non-metallic atoms playing a dominant role in the reemergence of superconducting and the influence of two Lifshitz transitions on superconducting properties.

**Introduction**

In unconventional superconductor systems, such as cuprates, iron-based superconductors, and heavy fermions superconductors, their superconducting transition temperatures ($T_c$) tend to demonstrate superconducting domes exactly as the composition and external conditions are transformed[1–5]. The emergence of superconductivity or at the superconducting transition temperature maximum is frequently accompanied by the intertwining of multiple symmetry-breaking orders such as charge density wave (CDW) order, electronic nematic order, and pair density wave (PDW) order, which exhibits a complex electronic phase diagram[6–9]. In this circumstance, understanding the micro-mechanism of pairing in high-temperature superconductors poses a significant challenge[2,10]. The Cu-O layer and Fe-As or Fe-Se layer in cuprate and iron-based superconductors have been shown important contributions to their superconductivity, respectively[11,12]. As a result, the interaction between layers plays a crucial role in high-temperature superconductors. Therefore, investigating a material with an uncomplicated electronic phase diagram and easily adjustable layered structure has significant implications for understanding the pairing mechanism of superconductivity.

Recently, natural van der Waals heterojunction materials have attracted extensive attention due to their interesting physical phenomena. For instance, chiral superconductors have been observed in $4H_b$-$TaS_2$ as well as topological nodal-point superconductivity due to the finite energy gap density of states (DOS) in the $1H$ layer[13,14]. In addition, there have been observations of double-layer superconductivity under high pressures in $4H_b$-$TaSe_2$[15]. The van der Waals heterojunction of these bulk materials provides an excellent platform to study interlayer interactions. Therefore, modulating the interlayer interaction of the material induces several intriguing physical phenomena, and pressure plays a crucial role in regulating interlayer interactions.

At ambient pressure, an extraordinary structure of $TaS_2$, which consists of six alternating stacked $1T$ (trigonal) and $1H$ (hexagonal) layers, and this unique arrangement belongs to the $R3m$ space group referred to as $6R$-$TaS_2$. The parent structure of $6R$-$TaS_2$, including $1T$- and $2H$-$TaS_2$ displays an extensive variety of properties at both ambient and high pressure[16,17], such as superconductivity and charge density wave (CDW)[18–

[23]. Therefore, the natural heterogeneous structure for 6$R$-TaS$_2$ should similarly exhibit a multitude of physical properties[16]. At ambient pressure, as the temperature gradually decreases, the structure of 6$R$-TaS$_2$ progressively destabilizes and the system is subject to structural distortion, which is realized as twelve tantalum atoms within the layer gradually move towards the central tantalum atom to the formation of David-star clusters of superlattices. The expansion of the cell leads to the folding of the Brillouin zone, modifying the electronic properties. Meanwhile, 6$R$-TaS$_2$ leads into a nearly commensurate charge density wave (NCCDW) phase at 320 K and a commensurate charge density wave (CCDW) transition at 305 K[16]. As the temperature reaches ~2.6 K, commencing occurs with the superconducting transition. The coexistence of superconductivity and CDW is revealed by the calculated electron-phonon coupling and low-temperature transmission electron microscopy (TEM), and the superconductivity is contributed by the $H$ layer whereas the $T$ layer is responsible for CDW, which is different from the previous TMDs[18–20,23–25], this allows us to obtain a deeper insight into how the interaction between layers influences superconductivity. For 6$R$-TaS$_2$, the layers are stacked by weak van der Waals force which is extremely susceptible to pressure modulation.

In this work, we utilize an in-situ method under high pressure to investigate the behavior of double-dome superconductivity and the reemergence of superconductivity in bulk 6$R$-TaS$_2$ polycrystalline. We discovered the first dome can be attributed to the enhancement of charge transfer between the $T$- and $H$-layer interlayer, which is associated with the shortens the distance of the interlayer. The formation of the second superconducting dome is primarily attributed to changes in the density of states resulting from Fermi surface reconstruction dominated by the $T$-layer. The superconducting reemergence of 6$R$-TaS$_2$ resulting from Lifshitz transition dominated by $p$-orbitals of S. These interesting findings provide a superior platform to investigate mechanisms of superconductivity.

## Results

## Double-dome and reemergence of superconductivity

In order to investigate the superconducting behavior of 6$R$-TaS$_2$ under high pressure, we performed the electric transport measurement without pressure-transmitting media (PTM) with the pressure range of 1.5 GPa~74 GPa. The results of electric transport measurements at various pressures are illustrated in Fig. 1 and Fig. 2, respectively. Fig. 1 shows the temperature dependence of resistance under different pressures. However, we have observed no signal associated with the CDW transition at the minimum pressure during the experiment in the whole temperature range, which suggests that the CDW order may be extremely susceptible to environments with hydrostatic pressure. Fig. 1a and Fig. 1d display the resistance curves $R$(T) of 6$R$-TaS$_2$ at a pressure range of 1.5–51.7 GPa.

Below 17.2 GPa, the resistance in the normal state gradually reduces as pressure increases. Beyond 17.2 GPa, the resistance of 6$R$-TaS$_2$ progressively increases, and exhibits the opposite trend with the low-pressure region. Moreover, the anomalous resistance can result from abnormal variations in carrier concentration and mobility[26], irregular variations of the Weyl point with pressure at the Fermi energy level[27], and boosted scattering of electron phonons at high pressures. However, considering the absence of topology in 6$R$-TaS$_2$, it is elevated resistances may be caused by abnormal variations in carrier concentration as well as improved electroacoustic coupling. Remarkably, the superconductivity of 6$R$-TaS$_2$ displays a complex behavior in the low-pressure region (1.5-32.5 GPa), manifesting a non-monotonic trend resembling the shape of an "M". At 1.5 GPa, the $T_c$ is first enhanced reaching about 5.6 K, which represents the first peak and is nearly double the superconducting transition temperature at ambient pressure. However, with the pressure further applied, $T_c$ gradually reduces to 4.7 K at 2.8 GPa resulting in the formation of the first superconducting dome, which has been extensively investigated in several TMDs systems[24,28]. Excitingly, $T_c$ has enhanced once again with pressure increasing and the second maximum value of $T_c$ occurred at 8.9 GPa reaching 5.5 K (Fig. 1b and Fig. 1c). However, while continuing further compression $T_c$ constantly shifts to low temperature, and the zero-resistance disappeared at 28.3 GPa. These intriguing findings imply that the superconducting properties of 6$R$-TaS$_2$ undergo enhancement through two distinct stages under pressure, indicating a highly intricate and challenging superconducting mechanism to comprehend. The superconducting behavior of 6$R$-TaS$_2$ under pressure displays a double dome-like shape, as observed in various materials like 4$H_b$-TaSe$_2$[15], K$_2$Mo$_3$As$_3$[29], and CsV$_3$Sb$_5$[30–33]. The explanation of the superconducting behavior of these materials in previous work can be attributed to interlayer coupling enhancement, interactions spanning several orbital energy levels, and variation of the CDW wave vector under pressure. At 35.3 GPa, the drop in resistance has been difficult to observe (Fig. 1e and Fig. 1f). And the resistance of pressurized in 6$R$-TaS$_2$ is almost invariant beyond 56.6 GPa.

Surprisingly, when the pressure is further increased to 56.6 GPa, we can see a new faint drop in resistance near 2 K, which indicates the sample of 6$R$-TaS$_2$ reemerges the superconducting state. The resistance decrease becomes more obvious with the increase of pressure and reaches zero resistance at 74 GPa which is the maximum pressure for the experiment as shown in Fig. 2a. To further verify whether the new faint drop in resistance is associated with a superconducting transition, we selected 71.1 GPa to measure the resistance of samples near $T_c$ under different magnetic field conditions. From Fig. 2b, it can be found that $T_c$ gradually decreases with the increase of the magnetic field, indicating the suppression of superconductivity. As a magnetic field of 0.2 T is applied to the sample, the resistance drop completely disappears. This provides experimental evidence for the superconducting reentrant of 6$R$-TaS$_2$.

To compare the SC-I and SC-II, we conducted an experiment where an external magnetic field was applied to the sample, as depicted in Fig. 2a and Supplementary Fig. S1. It is worth noting that the superconducting transition temperature, as a function of the magnetic field, exhibits a distinct positive curvature. This intriguing observation suggests that the widely used WHH equation, derived from the single-band model, is inadequate for accurately describing the behavior of 6$R$-TaS$_2$. Importantly, this trend is not limited to 6$R$-TaS$_2$ alone, as similar observations have been made in its parent compounds, namely 1$T$ and 2$H$[22,34]. In our experimental curve of $H_{c2}(0)$, we found that the critical field can be effectively described by an empirical equation, as illustrated in the accompanying insert (Supplementary Fig. S1). Remarkably, the estimated upper critical field for 6$R$-TaS$_2$ was determined to be 10.16 T for SC-I. Interestingly, the upper critical field of SC-I is much higher than that in SC-II, although neither of them has not exceeded the paramagnetic Pauli limit. In the SC-II, the binding force of Cooper pairs is much weaker compared to the SC-I. Later we will delve into the mechanism of complex superconductivity under various pressures.

**The structural analysis of the 6$R$-TaS$_2$**

In order to investigate whether the observed double dome-like superconductivity behavior and the reemergence of superconductivity in pressurized 6$R$-TaS$_2$ are related to the crystal structure phase transition under high pressure, we performed two runs of in situ high-pressure X-ray diffraction measurements at room temperature, as illustrated in Fig. 3 and Supplementary Fig. S2. The crystal structure of 6$R$-TaS$_2$ can be indexed with the $R$3m space group under ambient pressure as shown in Supplementary Fig. S3. The powder XRD pattern of 6$R$-TaS$_2$ under various pressures can be seen in Fig. 3a and Supplementary Fig. S2a, the samples exhibit a strong preferred orientation, in addition, all the XRD peaks move consecutively to higher angles, which means that the lattice parameter of 6$R$-TaS$_2$ is reduced under high pressure. Especially, peak <006> signifies a faster interlayer distance change compared to other peaks, reflecting the highly compressible nature of interlayer spacing in 6$R$-TaS$_2$ under pressure. Another piece of evidence is the significant reduction in the distance of the monolayer of $H$ and $T$ in 6$R$-TaS$_2$ (Supplementary Fig. S4), which is also one of the characteristics of quasi-2D materials[34,35]. Above 7.1 GPa, both the peak intensities of <110> and <116> are progressively deteriorating and disappear with further compression. The <006> peak consistently persists over the whole pressure range, indicating that it has not transformed from a quasi-2D to a quasi-3D structure at high pressures. In addition, no new peaks developed explain the 6$R$-TaS$_2$ stability up to 83.3

GPa (Fig. 3). The evolution of *d*-spacings as a function of pressure were depicted in Fig. 3b and Supplementary Fig. S2b. It is noteworthy that the value of <006> peak represents the distance of the interlayer that is more sensitive to pressure at low pressure. Above 2.8 GPa, XRD peaks shift constantly to higher angles. Supplementary Fig. S5 displays the full-profile refinement of the XRD pattern at 0.7 and 83.3 GPa, which indicates no structural transformation in 6*R*-TaS$_2$ happens under high pressure. This marks the first occurrence of superconducting reemergence without accompanying structural phase transitions among TMDs. Among the parent compounds of 6*R*-TaS$_2$ under high pressure, the superconductivity enhancement of 1*T*-TaS$_2$ is attributed to a layered to non-layered structural phase transition[34]; the emerging SC-II in 2*H*-TaS$_2$ is accompanied by structural change, and the enhancement of carrier density[22]. To further obtain the structural details of 6*R*-TaS$_2$ under high pressure, we fit the XRD patterns by the Rietveld method and extracted the volume of unit-cell and parameters as a function of pressure as described in Fig. 3c-3e and Supplementary Fig. S2c-2e. At 2.8 GPa, an inflection can be seen that constant *c* and *c*/*a* abruptly decrease and change in slope at the PV curve (Fig. 3d, e and Supplementary Fig. S2d, e). This observation indicates the occurrence of Fermi surface reconstruction in pressurized 6*R*-TaS$_2$, which is further supported by our resistance measurement of the first peak of $T_c$. Fig. 3c and Supplementary Fig. S2c show the results of fitting by Birch-

Murnaghan equation[36], generating the bulk modulus $B_0 = 35$ GPa and $V_0 = 350$ Å$^3$ at the low pressure region (0-3 GPa) and for high pressure region (4-83.3 GPa) yielding the $B_0 = 143.4$ GPa and $V_0 = 329$ Å$^3$. In pressurized 6$R$-TaS$_2$, the bulk modulus increases as pressure increases, showing that the material turns less compressible. This indicates that covalent bonding within the layers and layer coupling is improved as a result.

**Discussion**

There are many theories to account for the superconductivity transition temperature enhanced under high pressure: (1) Pressure improves the density of state near the Fermi energy enhanced electron-phonon coupling, leading to a rise of $T_c$. (2) Under high pressure, the structural instability leads to lattice softening enhancing the coupling strength of electrons and phonons. (3) The Fermi surface reconstruction namely Lifshitz transition also can enhance the $T_c$. As mentioned above, we have experimentally observed double-dome superconducting behavior as well as superconducting reemerge in pressurized 6$R$-TaS$_2$. However, it should be noted that these phenomena are independent of the structural phase transition. To acquire a better comprehension for the mechanism of the double dome and the reemergence of the superconductivity phenomenon under high pressure, we performed the first principal calculations on phonon dispersion and electronic properties, as shown in Fig. 4 and Supplementary Fig. S7-S10. Supplementary Fig. S6 exhibits the charge transfer of 6$R$-TaS$_2$ from the $T$- to $H$-layer. Below 5 GPa, the interlayer charge transfer dramatically increases associated with the first dome of superconductivity. The maximum value occurs at 15 GPa, but a sharp decrease in interlayer charge transfer is observed beyond 15 GPa.

Interestingly, our calculations of the band structure and the Fermi surface reveal two topological evolutions in the Fermi surface at 15 and 50 GPa. We plot the shape of the Fermi surface and projected band structure at 10, 15, and 20 GPa (Supplementary Fig. S7). A new hole-type pocket emerged at the Γ point at 15 GPa, indicating a potential Lifshitz transition, which was mainly contributed by the $T$-layer. However, the new hole-type pocket disappeared with increasing pressure, converging towards the valence band below the Fermi energy level at 20 GPa. This calculation result corresponds exactly to the second dome superconductivity of our electric transport experiment. Unexpectedly, with further compression, new hole-type pockets in the center of the Brillouin zone appeared at 50 GPa and gradually increased in size as the pressure increased to 60 GPa (Supplementary Fig. S8). The emergence of SC-II is attributed to the increase in density of states caused by Fermi surface reconstruction. To investigate the specific contributions of the $T$- or $H$-layer to the superconducting properties, we conducted electron-phonon coupling (EPC) calculations at 75 GPa. The phonon dispersion of 6$R$-TaS$_2$ at 75 GPa was illustrated in Supplementary Fig. S9, along with the phonon density of states contributed by the *T* and *H* layers, respectively. In the low-frequency region (< 220 cm$^{-1}$), the contributions of the *H* and *T* layers exhibit comparable magnitudes. However, in the high-frequency region (> 220 cm$^{-1}$), the *T* layer demonstrates a higher contribution compared to the *H* layer. To further elucidate this observation, we performed the Eliashberg spectral function $α^2F(ω)$ and the EPC constant $λ$, as presented in Fig. 4 (f). The $λ_{tot}$ is ~ 0.48, with 0.23 for the *H*-layer and 0.25 for the *T*-layer. It is obvious that the $λ$ of the *T*-layer is higher than that of the *H*-layer, and in combination with our XRD results, which indicate that the 6*R*-TaS$_2$ still exhibits quasi-two-dimensional properties, hence, we speculate that the SC-II originates from synergistic interaction between the *T*-layer and *H*-layer.

In the BCS theory regime, the $T_c$ and the DOS are proportional, therefore we meticulously analyze the evolution of the density of states under pressure, as shown in Fig. 4. At 60 GPa, the DOS at the Fermi level is presented by the *H*- and *T*-layers, respectively. Interestingly, the contribution of the Ta *d*-orbitals exhibits a peculiar behavior at the Fermi level, initially decreasing and subsequently increasing in a range of 40 to 70 GPa. In contrast, the *p*-orbitals of S experience a sharp increase, surpassing the contribution of the *d*-orbitals of Ta at 60 GPa. This unexpected trend suggests that the SC-II is predominantly governed by the *p*-orbitals of S. Further analysis, as depicted in Supplementary Fig. S10, reveals that the contributions from the *T* layer and the *p*-orbitals of S remain consistent from 40 to 70 GPa. Additionally, the density of states in the *H* layer starts to increase after 50 GPa. Consequently, the SC-II is primarily dominated by the *p*-orbitals of S atoms in both the *H* and *T* layers. This observation highlights the significance of the *p*-orbitals of S in driving the superconducting behavior, while the *d*-orbitals of Ta exhibit a more complex and subtle influence on the electronic properties.

**Conclusion**

Here, we study the double-dome superconductivity behavior of 6*R*-TaS$_2$ and the reemergence of superconducting under high pressure. XRD experiments have revealed that the exotic superconductivity is not related to any structural phase transition. In combination with DFT calculation, our researching suggests that the interlayer coupling changes the charge transfer between *T*- and *H*-layer 6*R*-TaS$_2$, thus forming the first dome-like superconducting. The second superconducting dome and the reemergence of superconductivity under pressure can be attributed to changes in the density of states resulting from Fermi surface reconstruction, in which the DOS of *T*-layer and S *p*-orbitals plays a crucial role. It unveils a highly intricate superconducting behavior resulting from two Lifshitz transitions. We have reported the first observation in TMDs that non-metallic atoms playing a dominant role in the reemergence of superconducting and the influence of two Lifshitz transitions on superconducting properties.

# Figures

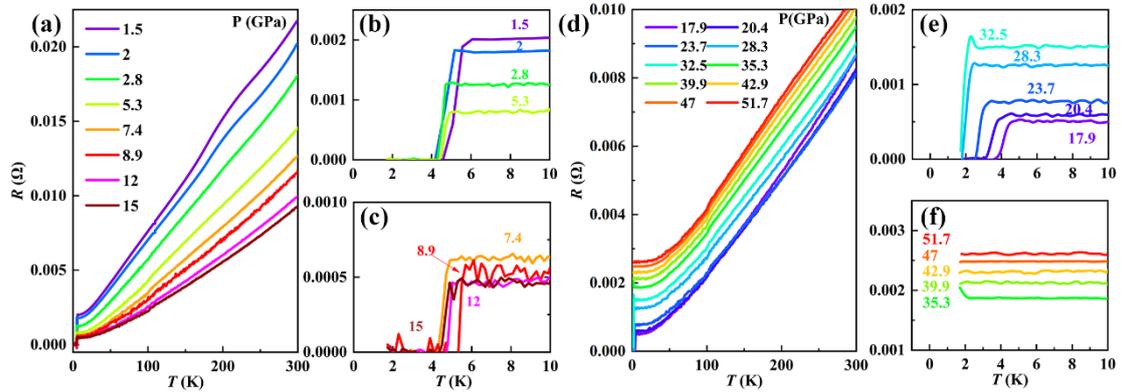

**FIG. 1 | Electronic transport properties as a function of pressure for 6*R*-TaS$_2$ under non-hydrostatic pressure conditions. (a)** Pressure dependence of resistance curves *R*(T) of 6*R*-TaS$_2$ at a pressure range of 1.5–15 GPa. **(b, c)** Enlarged view of Figure (a) at 0-10 K. **(d)** Pressure dependence of resistance curves *R*(T) of 6*R*-TaS$_2$ at a pressure range of 17.9–51.7 GPa. **(e, f)** Enlarged view of Figure (d) at 0-10 K.

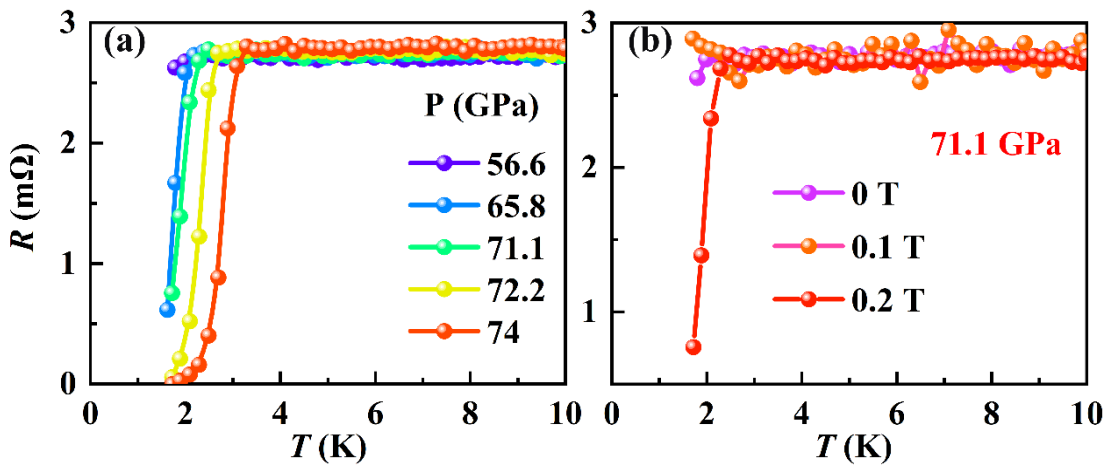

**FIG. 2 | (a)** Temperature dependence of resistance at 56.6 GPa to 74 GPa. **(b)** Temperature dependence of resistance of sample at various magnet fields up to 0.2 T.

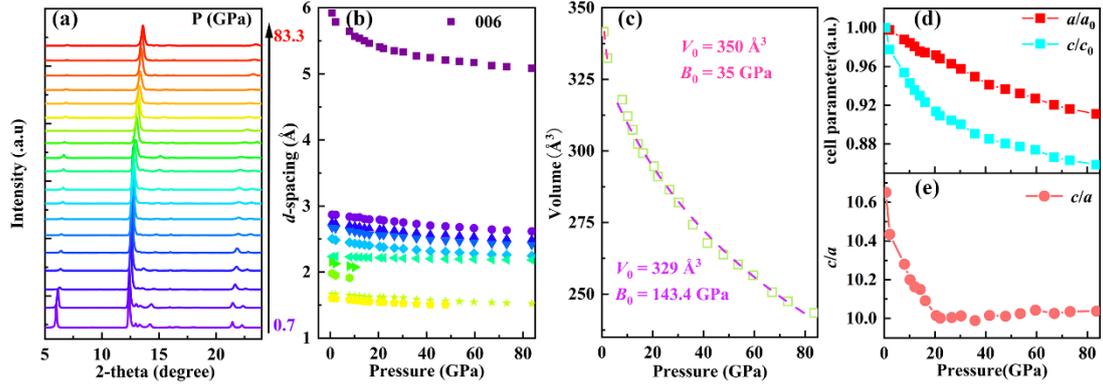

**FIG. 3 | Synchrotron XRD patterns of 6$R$-TaS$_2$ at room temperature under high pressure in run2 ($\lambda$ = 0.6199 Å). (a)** Representative XRD diffraction patterns under various pressures. **(b)** Variation of $d$-spacings under high pressure. **(c)** The pressure-dependent volume. The dashed line is the fitting result based on the Birch-Murnaghan equation of state. **(d, e)** Lattice parameters as a function of pressure $a/a_0$, $c/c_0$, and axis ratio $c/a$ which were extracted from Rietveld refinements.

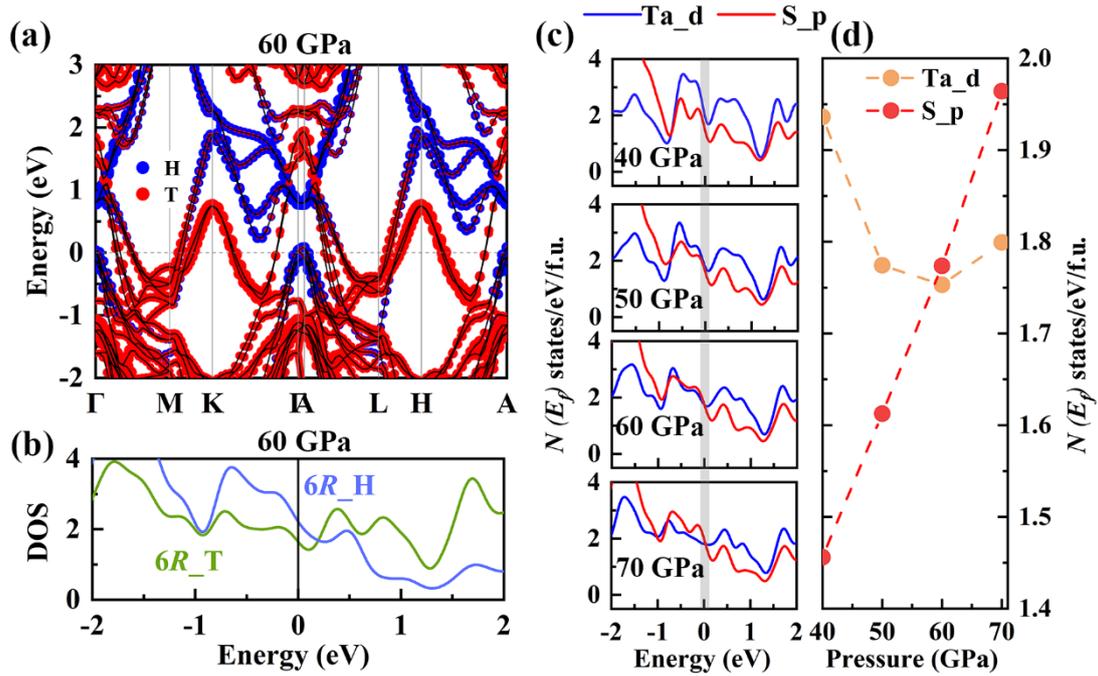

**FIG. 4 | The calculated band structure and DOS of 6*R*-TaS$_2$.** **(a)** Project band structure of the *T*- and *H*-layer at 60 GPa. The blue and red dots correspond to the *H* and *T* layers. The size of the dots represents the weight percentage. **(b)** Projected electronic DOS of *T*- and *H*-layer at 60 GPa. **(c)** Projected electronic DOS of Ta-*d* and S-*p* orbitals at different pressures. **(d)** The collected $N(E_f)$ of Ta-*d* and S-*p* orbitals under variation pressures at the Fermi level.

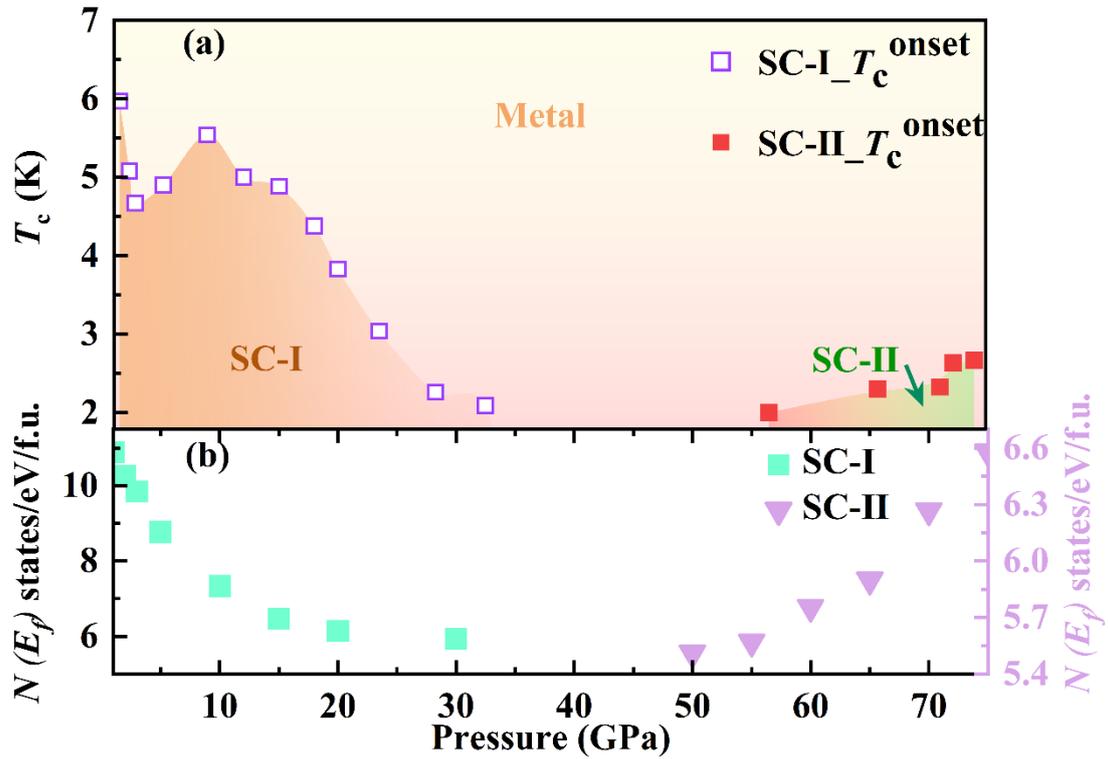

**Fig. 5 | (a)** Phase diagram of 6$R$-TaS$_2$ under high pressure. **(b)** The collected calculated electronic DOS under variation pressure at the Fermi level $N$(E$_f$) of 6$R$-TaS$_2$.

Supplementary Information

# Double Dome and Reemergence of Superconductivity in Pristine 6$R$-TaS$_2$ under Pressure

Xindeng Lv[1], Hao Song[1], Kun Chen[1], Sirui Liu[1], Yanping Huang[1,]*, Yuqiang Fang[3,]* and Tian Cui[1,2,]*

[1] Institute of High Pressure Physics, School of Physical Science and Technology, Ningbo University, Ningbo 315211, China

[2] State Key Laboratory of Superhard Materials, College of Physics, Jilin University, Changchun 130012, China

[3]State Key Laboratory of High Performance Ceramics and Superfine Microstructure, Shanghai Institute of Ceramics, Chinese Academy of Sciences, Shanghai, 200050, China

*Corresponding authors' e-mails: huangyanping@nbu.edu.cn; fangyuqiang@mail.sic.ac.cn; cuitian@nbu.edu.cn

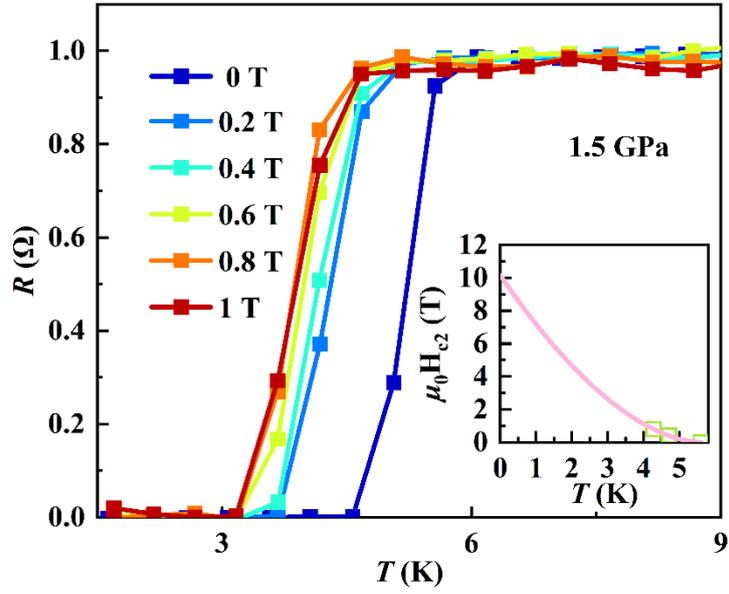

**FIG. S1.** Temperature dependence of resistance of sample at various magnet fields up to 1 T at 1.5 GPa. The $\mu_0 H_{c2}$ plots at different magnetic fields as shown in the insert. The solid line is the best fit of the equation $H_{c2}(T) = H_{c2}(0) * (1-T/T_c)^{1+\alpha}$.

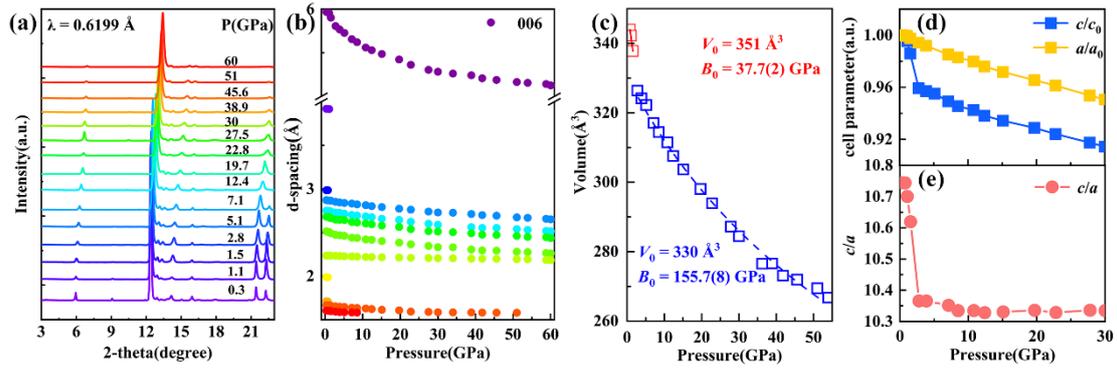

**FIG. S2**. Synchrotron XRD patterns of 6$R$-TaS$_2$ at room temperature under high pressure in run1 ($\lambda$ = 0.6199 Å). (a) Representative XRD diffraction patterns under various pressures. (b) Variation of $d$-spacings under high pressure. (c) The pressure-dependent volume. The dashed line is the fitting result based on the Birch-Murnaghan equation of state. (d, e) Lattice parameters $a/a_0$, $c/c_0$, and axis ratio $c/a$ as a function of pressure, which were extracted from Rietveld refinements.

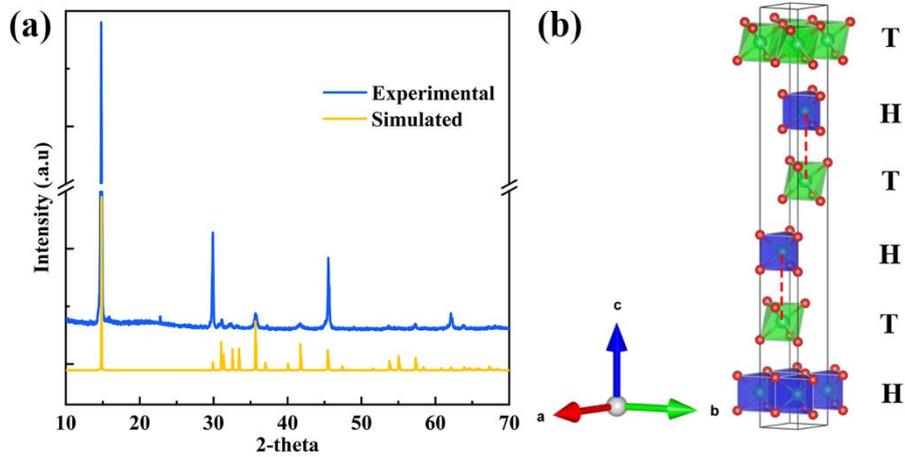

**FIG. S3.** (a) The polycrystal XRD patterns of 6$R$-TaS$_2$ at ambient pressure with wavelength of 1.5406 Å. (b) The crystal structure of 6$R$-TaS$_2$.

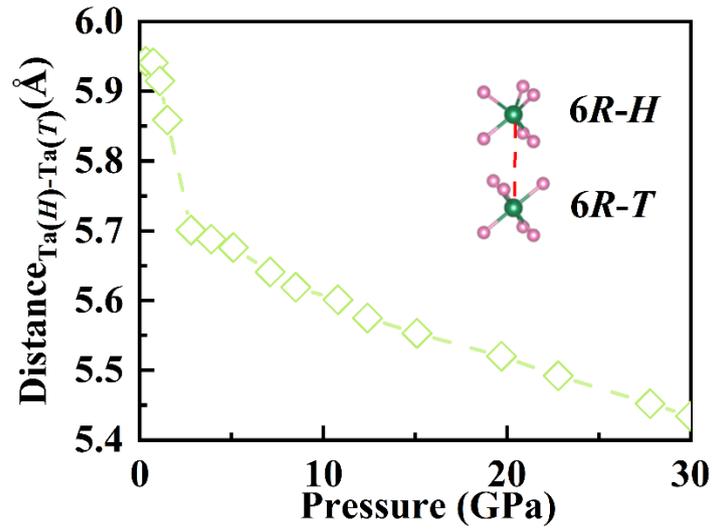

**FIG. S4.** The interlayer distance of 6$R$-TaS$_2$ as a function of pressure.

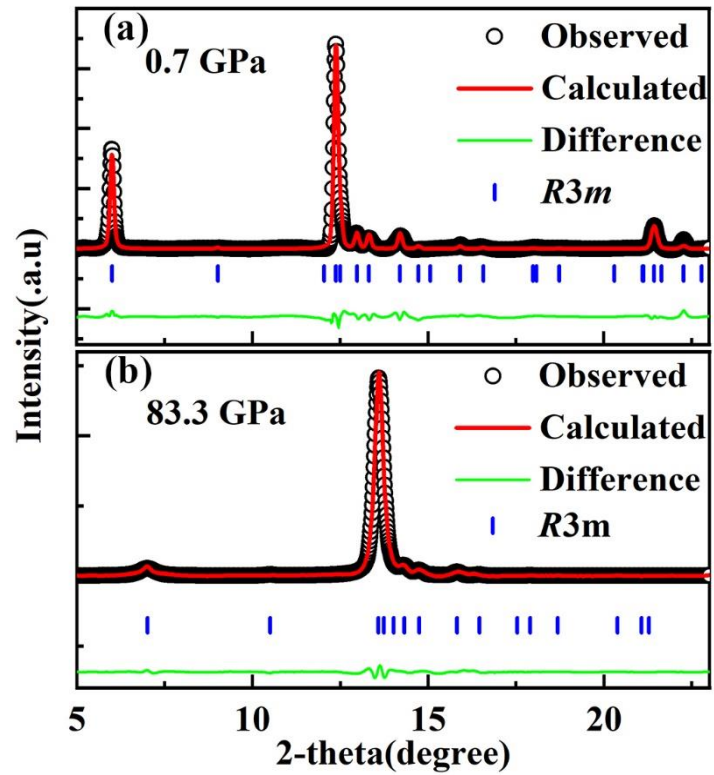

**FIG. S5.** Full-profile Rietveld refinements of XRD patterns in pressurized 6$R$-TaS$_2$ at 0.7 (a) and 83.3 (b) GPa.

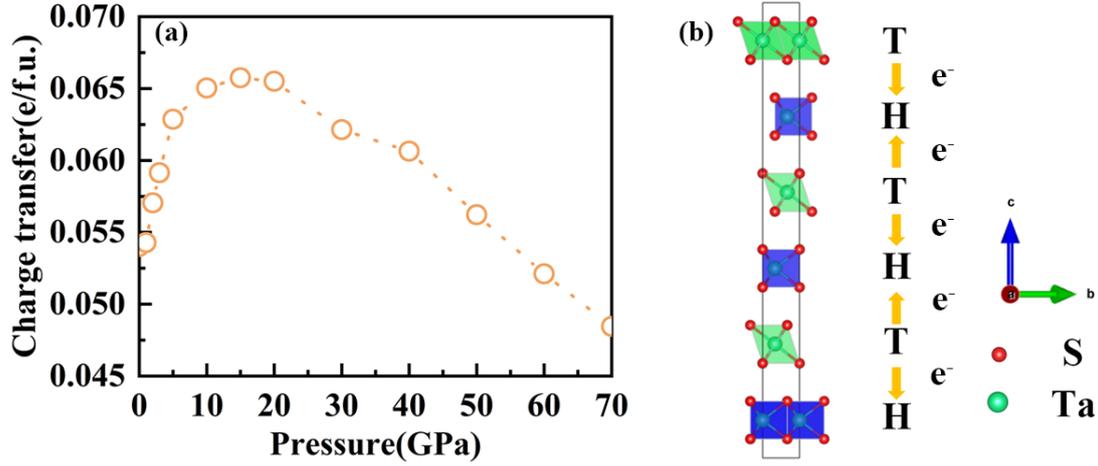

**FIG. S6**. (a) The calculated charge transfer of 6$R$-TaS$_2$ from $T$-layer to $H$-layer. (b) Schematic of interlayer charge transfer of 6$R$-TaS$_2$ from $T$-layer to $H$-layer.

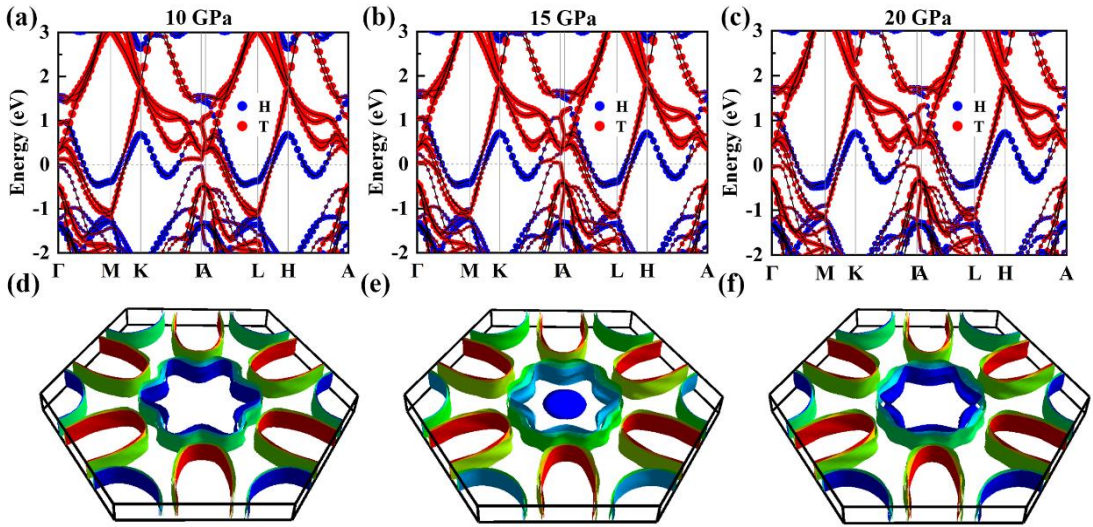

**FIG. S7**. The calculated band structure and Fermi surface of 6$R$-TaS$_2$. (a-c) Projected band structure at 10 GPa, 15 GPa, and 20 GPa, respectively. The blue and red dots correspond to the $H$ and $T$ layers. The size of the dots represents the weight percentage. (d-f) Fermi surface of 6$R$-TaS$_2$ at 10 GPa, 15 GPa, and 20 GPa.

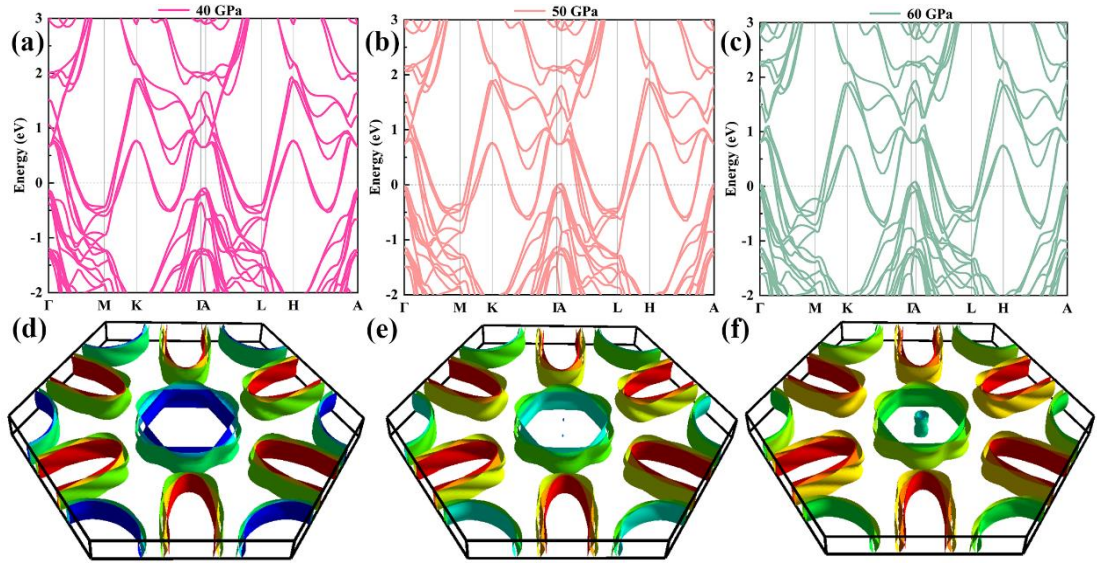

**FIG. S8.** The calculated band structure and Fermi surface of 6R-TaS$_2$. (a-c) Band structure of 6R-TaS$_2$ at 40 GPa, 50 GPa, and 60 GPa. (d-f) Fermi surface of 6R-TaS$_2$ at at 40 GPa, 50 GPa and 60 GPa.

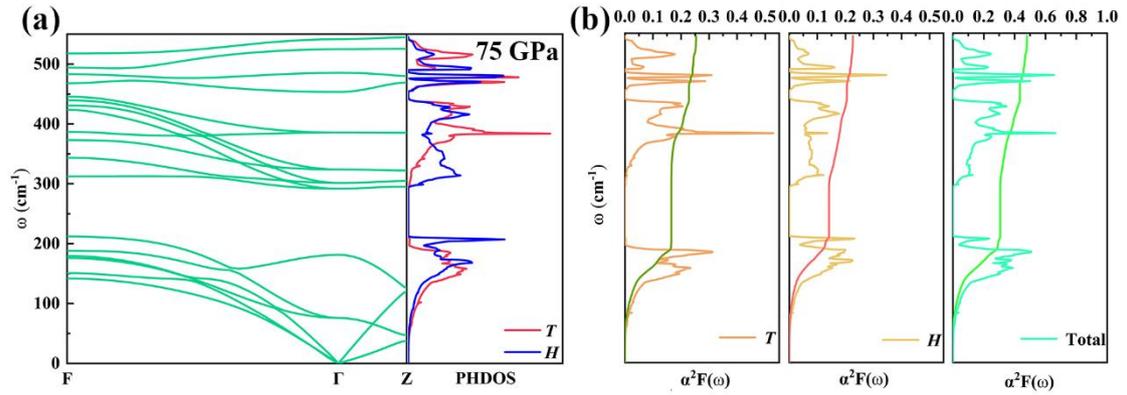

**FIG. S9.** The calculated phonon dispersion and EPC of 6R-TaS$_2$. (a) The phonon dispersion and PHDOS of 6R-TaS$_2$ at 75 GPa. (b) Eliashberg spectral function $\alpha^2F(\omega)$ and the EPC constant $\lambda$ are projected to the T- and H-layers at 75 GPa.

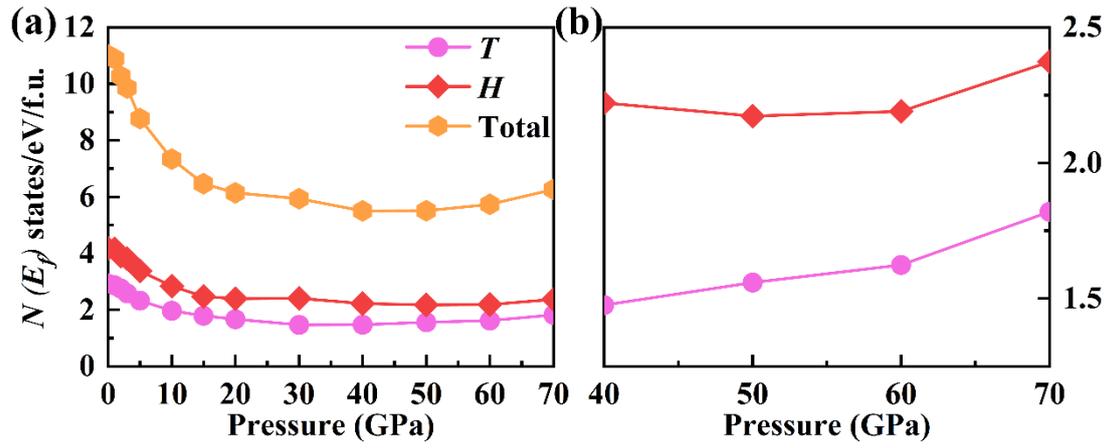

**FIG. S10.** (a) The collected DOS projected to the *T*- and *H*-layer of 6*R*-TaS$_2$ under different pressures. (b) Enlarged view of Figure (a) at 40-70 GPa.